\documentclass[prl,twocolumn,superscriptaddress,longbibliography,notitlepage]{revtex4-1}
\usepackage{amsmath,amssymb,amsthm,amsfonts,mathrsfs}
\usepackage{bm}
\usepackage{textcomp}
\usepackage{graphicx}
\usepackage{color}
\usepackage{tikz,pgf}%
\usetikzlibrary{shadings}
\usepackage{dsfont}
\usepackage{fontenc}
\usepackage{amsfonts}
\usepackage{booktabs}
\usepackage[colorlinks=true, letterpaper=true, pdfstartview=FitV, linkcolor=blue, citecolor=blue, urlcolor=blue]{hyperref}

\begin{document}

\title{Two-dimensional second-order topological insulator in graphdiyne}

\author{Xian-Lei Sheng}
\email{xlsheng@buaa.edu.cn}
\affiliation{School of Physics, and Key Laboratory of Micro-nano Measurement-Manipulation and Physics, Beihang University, Beijing 100191, China}
\address{Research Laboratory for Quantum Materials, Singapore University of Technology and Design, Singapore 487372, Singapore}

\author{Cong Chen}
\affiliation{School of Physics, and Key Laboratory of Micro-nano Measurement-Manipulation and Physics, Beihang University, Beijing 100191, China}
\address{Research Laboratory for Quantum Materials, Singapore University of Technology and Design, Singapore 487372, Singapore}

\author{Huiying Liu}
\address{Research Laboratory for Quantum Materials, Singapore University of Technology and Design, Singapore 487372, Singapore}

\author{Ziyu Chen}
\affiliation{School of Physics, and Key Laboratory of Micro-nano Measurement-Manipulation and Physics, Beihang University, Beijing 100191, China}

\author{Zhi-Ming Yu}
\email{zhiming\_yu@sutd.edu.sg}
\address{Research Laboratory for Quantum Materials, Singapore University of Technology and Design, Singapore 487372, Singapore}
\affiliation{Key Lab of Advanced Optoelectronic Quantum Architecture and Measurement (MOE), Beijing Key Lab of Nanophotonics $\&$ Ultrafine Optoelectronic Systems, and School of Physics, Beijing Institute of Technology, Beijing 100081, China}

\author{Y. X. Zhao}
\email{zhaoyx@nju.edu.cn}
\address{National Laboratory of Solid State Microstructures and Department of Physics, Nanjing University, Nanjing 210093, China}
\address{Collaborative Innovation Center of Advanced Microstructures, Nanjing University, Nanjing 210093, China}

\author{Shengyuan A. Yang}
\address{Research Laboratory for Quantum Materials, Singapore University of Technology and Design, Singapore 487372, Singapore}

\begin{abstract}
A second-order topological insulator (SOTI) in $d$ spatial dimensions features topologically protected gapless states at its $(d-2)$-dimensional boundary at the intersection of two crystal faces, but is gapped otherwise. As a novel topological state, it has been attracting great interest, but it remains a challenge to identify a realistic SOTI material in two dimensions (2D).
Here, based on combined first-principles calculations and theoretical analysis, we reveal the already experimentally synthesized 2D material graphdiyne as the first realistic example of a 2D SOTI, with topologically protected 0D corner states. The role of crystalline symmetry, the robustness against symmetry-breaking, and the possible experimental characterization are discussed. Our results uncover a hidden topological character of graphdiyne and
promote it as a concrete material platform for exploring the intriguing physics of higher-order topological phases.
\end{abstract}

\maketitle

{\color{blue}{\em Introduction.}}---The discovery of topological insulators (TIs) has generated a vast research field~\cite{Hasan_RMP,Qi_RMP,Bansil_RMP,ShunQingShen_TI}. Often considered as its defining property, a TI in $d$ spatial dimensions has an insulating bulk, but features protected gapless states on its $(d-1)$-dimensional boundaries. Recently, the notion was extended to a new class of topological phases, known as higher-order TIs~\cite{ZhangFan_PRL2013,Hughes2017,Langbehn2017,SongZD2017,Hughes2017b,Schindler2018SA}. An $n$-th order TI features protected gapless states at its $(d-n)$-dimensional boundary at the intersection of $n$ crystal faces, but is gapped otherwise. For example, a second-order topological insulator (SOTI) in 2D (3D) hosts topological states located at its 0D corners (1D hinges) between distinct edges (surfaces) which are gapped. This indicates that the $(d-1)$-dimensional boundary of a SOTI is itself an insulator with topological classifications. So far, higher-order TIs have been proposed in a few 3D materials~\cite{Schindler2018,Schindler2018SA,Yue2019ws,ZhangFan_PRL2013,WangZhijun_arXiv,Bradlyn2017,Zhang2019tp,Vergniory2019ub,TangNP,TangNature,XuYF2019} and in some artificial systems~\cite{Kruthoff2017,Ezawa2018b,Ezawa2018L,Imhof2018wj,SerraGarcia2018,Peterson2018,KariyadoSR,Yuhan_arXiv,Xue2019to}. However, possibly due to the less number of 2D materials and the difficulty in characterizing the higher-order phases, a realistic 2D SOTI material has not been found yet, which poses a great challenge for the research on higher-order TIs.

Meanwhile, in the field of 2D materials, a new carbon allotrope with single-atom thickness---the graphdiyne (GDY)---has been attracting significant interest. Originally predicted in 1987~\cite{GDY1987}, the material was synthesized in experiment by Li and co-workers in 2010 through an in-situ cross-coupling method~\cite{LiYL2010}. Since then, a variety of synthesis methods for GDY have been developed, and its properties have been actively explored (see \cite{Ruben2017,Jia2017cs,YulinagLi_ChemRev,ZhangJin_ChemSocRev} and references therein), demonstrating its potential applications in environmental science~\cite{GaoX2016am}, energy~\cite{Lu2018ur,Hui2018vu}, catalysis~\cite{Yang2013cm,WangD2014,Ren2015,Xue2018vq}, and electronics~\cite{Li2015nc,ShuaiZG2013jpcl,Jiao2011cc}. Notably, GDY is a semiconductor with a bandgap $\sim 0.5$ eV~\cite{Long2011fu,Narita1998}. Its topological properties have not been carefully investigated before, because owing to its negligible spin-orbit coupling (SOC), the material must be topologically trivial according to the \emph{conventional} classification of time-reversal invariant insulators. However, this argument does not forbid a higher-order topology.

In this work, we theoretically predict GDY as the first realistic example of a 2D SOTI. We show that the 2D bulk of GDY features a band inversion at the $\Gamma$ point between two doublet states. As a result, a pair of gapped edge bands appear on its typical 1D edges, which are captured by the 1D Dirac model with a mass term. Such an edge spectrum admits a $\mathbb{Z}_2$ classification, so protected corner state must arise as the topological domain-wall state at the intersection between two edges if they belong to different topological classes. We show that this is the case for two edges related by a mirror symmetry, and hence for a hexagonal-shaped GDY nanodisk with neighboring edges related by these mirrors, there will be six protected states localized at the six corners, with energy pinned in the bulk gap due to an approximate chiral (sublattice) symmetry. Furthermore, we demonstrate that the exact crystalline symmetries (such as the mirror) are not essential: the 0D topological boundary states are robust against symmetry-breaking perturbations and shape imperfections. This will facilitate the experimental characterization of the SOTI phase in GDY.
\begin{figure}
  \includegraphics[width=8.5cm]{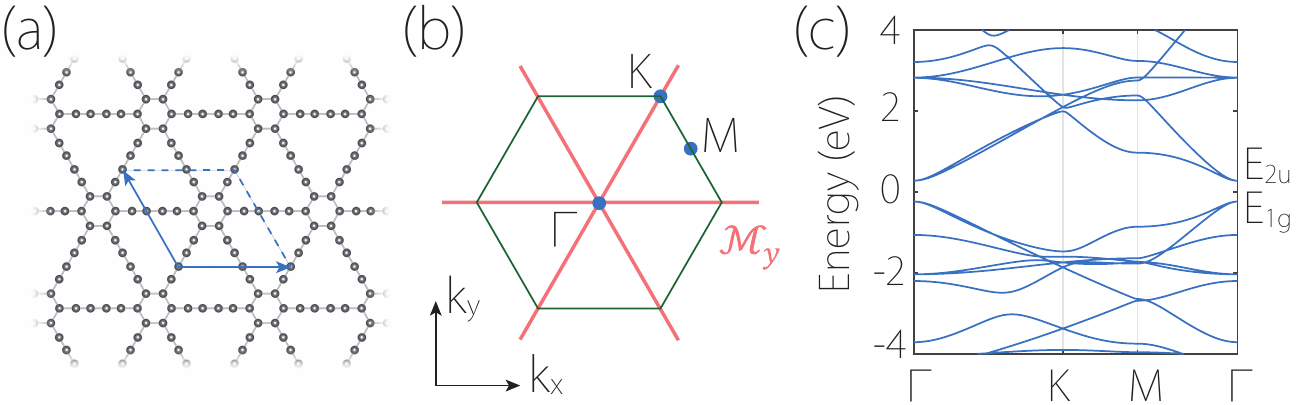}
  \caption{(a) Crystal structure of GDY. (b) shows the Brillouin zone. The red lines indicate the three equivalent mirrors $\mathcal{M}_y$. (c) Bulk electronic band structure of GDY. The CBM and the VBM states have $E_{2u}$ and $E_{1g}$ symmetry characters, respectively.}
\label{crystal}
\end{figure}

{\color{blue}{\em Lattice structure.}}---GDY is a carbon allotrope with a 2D planar network structure [Fig.~\ref{crystal}(a)], which may be viewed as formed by inserting the diacetylenic linkage between two neighboring benzene rings in a graphene structure. The lattice is completely flat, with a single-atom thickness, and has the same $p$6m symmetry as graphene.
Consisting of both $sp$- and $sp^2$-hybridized carbons, GDY exhibits a high $\pi$-conjugation, which helps to stabilize the planar structure and lower the system energy.
It has been found that GDY is the most stable non-natural carbon allotrope containing the diacetylene bonds~\cite{Haley1997}. The lattice constant obtained from our first-principles calculations is 9.46~\AA~(see SM~\cite{SM}), which is consistent with previous results~\cite{Long2011fu,Narita1998}.

{\color{blue}{\em Bulk band structure.}}---Figure~\ref{crystal}(c) shows our calculated band structure of GDY. Note that since SOC effect is negligibly small, it is neglected in the calculation, and the system can be effectively treated as spinless in the following analysis.

From Fig.~\ref{crystal}(c), one observes that GDY is a direct bandgap semiconductor with gap $\sim 0.51$ eV, in agreement with previous studies~\cite{Long2011fu,Narita1998}. (The bandgap is increased to $\sim 1.10$ eV with GW approach~\cite{LuoGF2011}, while the band topology remains the same. See the Supplemental Material~\cite{SM}.) The direct gap occurs at $\Gamma$ of the Brillouin zone (BZ). Notably, both the conduction band minimum (CBM) and the valence band maximum (VBM) are formed by degenerate doublet states. The CBM doublet corresponds to the two-dimensional irreducible representation $E_{2u}$ of the $D_{6h}$ group; while the VBM corresponds to the $E_{1g}$ representation. The two doublets have opposite parities under inversion.

Owing to the preserved time reversal symmetry $\mathcal{T}$ and the negligible SOC, the system is trivial according to the conventional
characterization of 2D TIs.
However, we find that the band edge configuration for GDY actually indicates an inverted band ordering at $\Gamma$. Indeed, in the atomic insulator limit, which can be achieved, e.g., by expanding the lattice, the $E_{2u}$ doublet is found to be energetically below $E_{1g}$. This band inversion can also be verified by comparing the parity eigenvalues of the occupied bands at the four inversion-invariant momenta, including $\Gamma$ and the three $M$ points. For GDY, the three $M$ points are equivalent, so we only need to compare $\Gamma$ with one $M$. Let $n_{+}^{k_i}$ ($n_{-}^{k_i}$) denote the number of occupied bands with positive (negative) parity eigenvalue at $k_i$. Over the total 36 valence bands considered in the calculation, we find that at $M$, $n^M_+=18$, and $n_-^M=18$; whereas at $\Gamma$, $n^\Gamma_+=20$, and $n_-^\Gamma=16$. Evidently, the difference $(n_+^\Gamma-n_+^M)=2$ corresponds to the band inversion between $E_{1g}$ and $E_{2u}$ at $\Gamma$. This band inversion plays a key role in the existence of corner states, as we shall see below.

\begin{figure}[t!]
\includegraphics[width=8.5 cm]{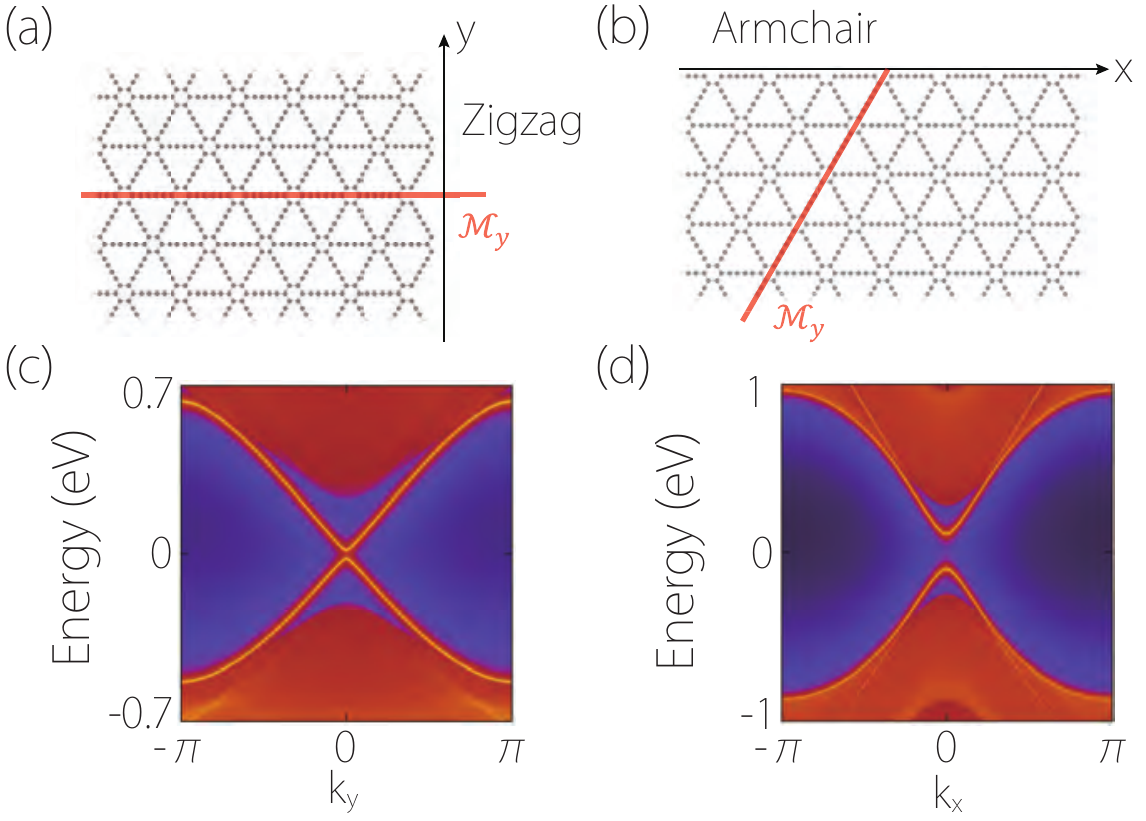}
\caption{(a) Zigzag and (b) armchair edge of a semi-infinite sheet of GDY. $\mathcal{M}_y$ is preserved for the zigzag edge, but not for the armchair edge. (c) and (d) show the projected spectra for  zigzag edge and  armchair edge, respectively. Note that there is a small gap between the edge bands in (c). }
\label{surf}
\end{figure}

{\color{blue}{\em Edge spectra and corner states.}}---A trivial insulator is usually not expected to have edge states inside the bulk gap. However, we find the existence of in-gap edge states for typical edges of GDY. In Figs.~\ref{surf}(c) and \ref{surf}(d), we plot the edge spectra for the zigzag and the armchair edges illustrated in Figs.~\ref{surf}(a) and \ref{surf}(b). One can observe the following features. First, there are two edge bands appearing on each edge (the bright curves). Second, these edge bands are \emph{not} gapless, which is consistent with the bulk not being a conventional TI. Nevertheless, one notices the edge-band gap for the zigzag edge is quite small ($\sim 0.0385$ eV), such that the spectrum in Fig.~\ref{surf}(c) mimics that for a quantum spin Hall insulator. Meanwhile, the gap for the armchair edge is relatively large $\sim 0.21$ eV. Later, we shall see that the presence of these edge bands is connected with the bulk band inversion.

\begin{figure}[t!]
\includegraphics[width=8.5 cm]{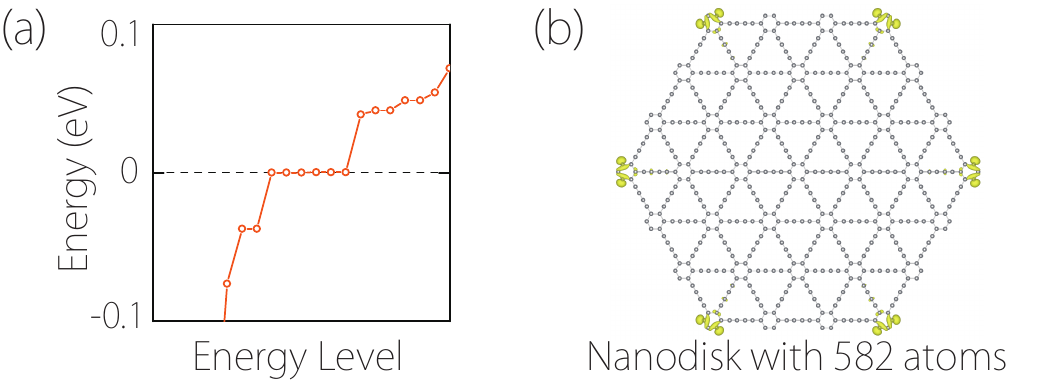}
\caption{(a) Energy spectrum of the hexagonal-shaped GDY nanodisk shown in (b). Energy levels are plotted in ascending order. (b) also shows the charge distribution of the six zero-energy states, which are localized at corners.}
\label{zeromode}
\end{figure}

Next, we explore the corner states, which is the hallmark of a 2D SOTI. To this end, we calculate the spectrum for a 0D geometry, namely, a GDY nanodisk. For concreteness, we take a hexagonal-shaped nanodisk [see Fig.~\ref{zeromode}(b)], which corresponds to the geometry obtained
in experiment from the bottom-up interfacial synthesis approach~\cite{MatsuokaJACS}.
The obtained discrete spectrum for the nanodisk is plotted in Fig.~\ref{zeromode}(a). Remarkably, one observes six states degenerate at zero energy, i.e., the Fermi level. {The spatial distribution of these zero-energy modes can be visualized from their charge distribution, as shown in Fig.~\ref{zeromode}(b).} Clearly, these states are well localized at the six corners, so they correspond to isolated corner states. (Finite-size effects can split the degeneracy, but the splitting is exponentially suppressed with the disk size.) At exact half filling, three of the six states will be occupied, and the excitation of the system becomes gapless, with zero-energy corner excitations.

{\color{blue}{\em Topological origin of the corner states.}}---To demonstrate the topological origin of the corner states, we construct an edge theory, and show that the nanodisk edges have a $\mathbb{Z}_2$ topological classification, then a corner state will arise as the topological domain-wall state at the intersection of two edges belonging to distinct classes.

We start with a model capturing the bulk low-energy band structure, which is around $\Gamma$ and features a band inversion. Hence, we construct a $k\cdot p$ model expanded around $\Gamma$, subjected to the $D_{6h}$ and $\mathcal{T}$ symmetry. The generators for $D_{6h}$ can be chosen as the threefold rotation $C_{3z}$, the twofold rotation $C_{2z}$ (equivalent to the inversion $\mathcal{P}$ for the 2D spinless case), and the vertical mirror $\mathcal{M}_y$ perpendicular to $y$.
In the basis of the two doublets $(E_{1g},E_{2u})^T$, they are represented by
\begin{equation}
C_{3z}=e^{-i\frac{\pi}{3}s_y}\tau_0, \ \   C_{2z}=\mathcal{P}=s_0\tau_z,\ \  \mathcal{M}_y=s_z\tau_0,
\end{equation}
where the Pauli matrices $s$ and $\tau$ represent two \emph{pseudospin} degrees of freedom, $\tau$ acts on the two doublets, and $s$ acts on the two degenerate states within each doublet, $s_0$ and $\tau_0$ are the $2\times 2$ identity matrix.
In addition, $\mathcal{T}=\mathcal{K}$, with $\mathcal{K}$ the complex conjugation, as for the spinless case.

Constrained by these symmetries, the bulk model expanded to $k$-quadratic order reads
\begin{eqnarray}\label{H2D}
\mathcal{H}_\text{2D}({\bm{k}})&=&W-(m_0-m_1k^2)\tau_z+v(k_x s_z-k_y s_x)\tau_y \nonumber\\
&&+\big[(k_x^2-k_y^2)s_z+2k_xk_y s_x\big](c_1\tau_0+c_2\tau_z).
\end{eqnarray}
where $W=w_0+w_1k^2$, $k=|\bm k|$; $w_i$, $m_i$, $c_i$, and $v$ are real parameters. We have $m_1>0$, because for the trivial vacuum (at $k\rightarrow \infty$) $E_{1g}$ is above $E_{2u}$, as we have discussed. Whether a band inversion occurs or not is determined by the sign of $m_0$; for GDY, the band inversion is signaled by $m_0>0$. From the spectra in Fig.~\ref{crystal}(c) and Fig.~\ref{surf}, one also observes that the system has an \emph{approximate} chiral symmetry, namely, the low-energy spectra are roughly symmetric about zero-energy. It derives from a sublattice symmetry of the structure~\cite{SM}. Such an approximate chiral symmetry often emerges in carbon allotropes~\cite{Chenyuanping_Nanolett,SAY_NodalSurface}. { In model (\ref{H2D}) [and the following model (\ref{Eq5})], the chiral symmetry is represented as $\mathcal{C}=\tau_x$,} such that when the first term and the $c_1$ term can be neglected, we have $\{\mathcal{C},\mathcal{H}_\text{2D}\}=0$.

To derive the edge theory, one can directly solve the edge spectrum for a given edge orientation, but here, instead, we proceed with a more intuitive argument by utilizing the mirror symmetry $\mathcal{M}_y$, following the approach by Langbehn \emph{et al.}~\cite{Langbehn2017}. This also helps to explain the edge spectrum observed in Fig.~\ref{surf}(c).

Let's consider the $\mathcal{M}_y$-invariant path $k_y=0$ in the bulk BZ. On this path, the two mirror subspaces with opposite $\mathcal{M}_y$ eigenvalues $\pm$ are decoupled. For each mirror subspace, one can evaluate its Berry phase for all occupied (valence) bands~\cite{ZhangPRL2013CL}, given by
\begin{equation}
\gamma_\pm=\oint_{k_y=0}\text{Tr}[\mathcal{A}_{\pm}(\bm k)]\cdot d\bm k,
\end{equation}
where $\mathcal{A}_{\pm}(\bm k)$ is the non-Abelian Berry connection for the occupied bands in the mirror subspace $\pm$.
This Berry phase must be quantized (in units of $\pi$) due to the $\mathcal{P}$ and $\mathcal{T}$ symmetries, and it represents the electric polarization for a mirror subspace of the 1D system with $k_y=0$~\cite{Smith_PRB}.
Connected with the band inversion at $\Gamma$, calculations both from the effective model and from the first-principles approach show that $\gamma_+=\gamma_-=\pi$, dictating the presence of one edge state for each mirror subspace at an edge where the $k_y=0$ path has a finite projection. Particularly, on the zigzag edge in Fig.~\ref{surf}(a) which preserves $\mathcal{M}_y$, the two mirror subspaces and hence the two edge states at the center of the edge BZ are still decoupled, such that each edge state will be pinned at zero energy as required by the chiral symmetry. This explains the spectrum observed in Fig.~\ref{surf}(c), where the small gap is because $\mathcal{C}$ is not an exact symmetry for the system. Meanwhile, for the armchair edge, which does not preserve $\mathcal{M}_y$, the two edge states would generally repel each other from zero energy. This explains the spectrum in Fig.~\ref{surf}(d).

To have a quantitative description, we derive an edge model from the bulk $\mathcal{H}_\text{2D}$. For simplicity, let's turn off the quadratic terms, as they will not affect the essential physics. First, consider a flat edge at $x=0$ that preserves $\mathcal{M}_y$ (with the zigzag edge in mind), with GDY occupying $x<0$. The edge states are solved from
\begin{equation}
  \tilde{\mathcal{H}}\psi=E\psi,
\end{equation}
with
\begin{equation}\label{Eq5}
\tilde{\mathcal{H}}=m(x)\tau_z+v (-is_z\partial_x- k_y s_x)\tau_y,
\end{equation}
where $m(x<0)=-m_0$, and $m(x>0)=+M$ with a large $M\rightarrow+\infty$ for the vacuum side  \cite{Zhang2012PRB}. At $k_y=0$, the states with $s_z=\pm 1$, i.e., with opposite $\mathcal{M}_y$ eigenvalues, are decoupled, consistent with our previous analysis.  Considering $s_z=+1$, the equation is reduced to a Jackiw-Rebbi problem~\cite{Jackiw_PRD}, with a topological zero-energy edge mode
\begin{equation}
\psi_+=\frac{1}{A}\left(
                  \begin{array}{c}
                    1 \\
                    0 \\
                  \end{array}
                \right)_s\otimes\left(
                  \begin{array}{c}
                    1 \\
                    -1 \\
                  \end{array}
                \right)_\tau\exp\left[-\frac{1}{v}\int^x_0\ m(x')dx'\right],
\end{equation}
where $A$ is a normalization factor. Similarly, the other zero-mode for $s_z=-1$ is given by
\begin{equation}
  \psi_-=\frac{1}{A}\left(
                  \begin{array}{c}
                    0 \\
                    1 \\
                  \end{array}
                \right)_s\otimes\left(
                  \begin{array}{c}
                    1 \\
                    1 \\
                  \end{array}
                \right)_\tau\exp\left[-\frac{1}{v}\int^x_0\ m(x')dx'\right].
\end{equation}
Now, expanded at $k_y=0$ on the basis of these two states, the 1D edge model is given by
\begin{equation}\label{Hedge}
  \mathcal{H}_\text{edge}(k)=vk \sigma_y,
\end{equation}
where $k$ is the wave vector along the edge, and the Pauli matrices $\sigma$ act on the space of $(\psi_-,\psi_+)^T$. Within this reduced edge space, the symmetry operators are represented by
\begin{equation}
\mathcal{M}_y=-\sigma_z,\qquad \mathcal{C}=\sigma_z.
\end{equation}
One observes that $\mathcal{M}_y$ and $\mathcal{C}$ forbid any mass term for the edge model (\ref{Hedge})~\cite{Koshino2014}. This Dirac edge model consistently describes the zigzag edge spectrum in Fig.~\ref{surf}(c) {(the high-order terms in $\mathcal{H}_\text{2D}$ may weakly break the chiral symmetry and open a small gap, but they will not qualitatively change the Dirac type spectrum)}.

\begin{figure}[t!]
\includegraphics[width=8.5 cm]{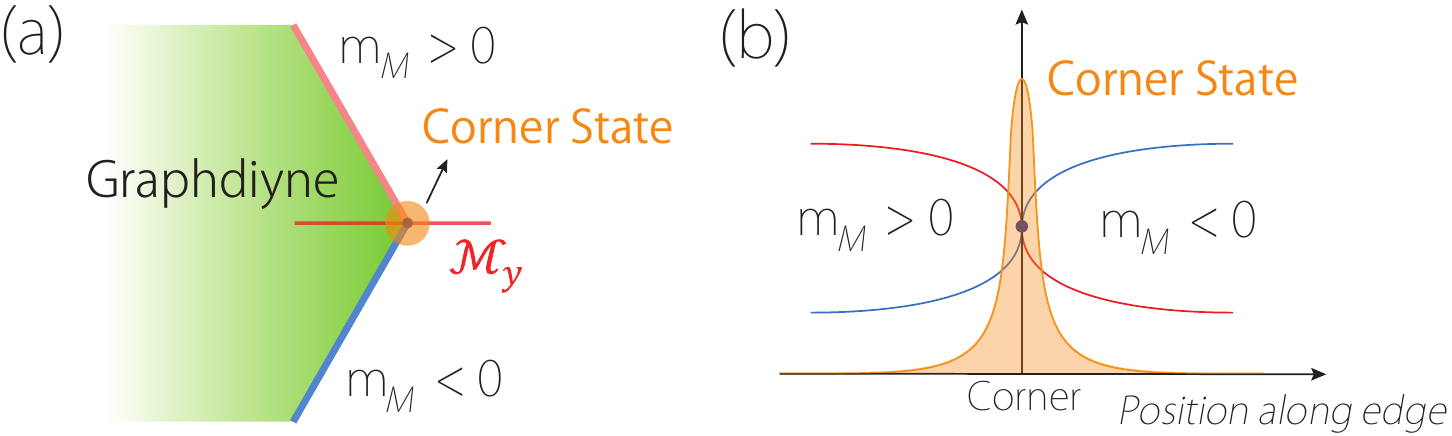}
\caption{(a) Schematic figure showing that two edges related by $\mathcal{M}_y$ must have opposite Dirac mass hence distinct topological classification. (b) Therefore, the corner at the intersection of the two edges must host a protected corner state as a topological domain-wall state.}
\label{model}
\end{figure}

For an edge that does not preserve $\mathcal{M}_y$ (such as the armchair edge), the edge model $\mathcal{H}_\text{edge}$ will generally be gapped by the mass term
\begin{equation}
  \Delta_M=m_M\sigma_x.
\end{equation}
It is well known that the sign of the mass term $\text{sgn}(m_M)$ gives a $\mathbb{Z}_2$ topological classification of the 1D Dirac model~\cite{ShunQingShen_TI}.
Thus, protected 0D corner mode must exist at the intersection between two edges belonging to distinct topological classes~\cite{Jackiw_PRD,ZhangF2018,YanZB2018}. Particularly, because this mass term is odd under $\mathcal{M}_y$, two edges connected by $\mathcal{M}_y$ must have opposite masses.
We therefore have a \emph{sufficient} condition: A protected 0D mode must exist at the corner where two $\mathcal{M}_y$-related topological edges meet, regardless of the detailed edge geometry. This is illustrated in Fig.~\ref{model}, and is consistent with the first-principles result for the nanodisk in Fig.~\ref{zeromode}.

\begin{figure}[t!]
\includegraphics[width=8.5 cm]{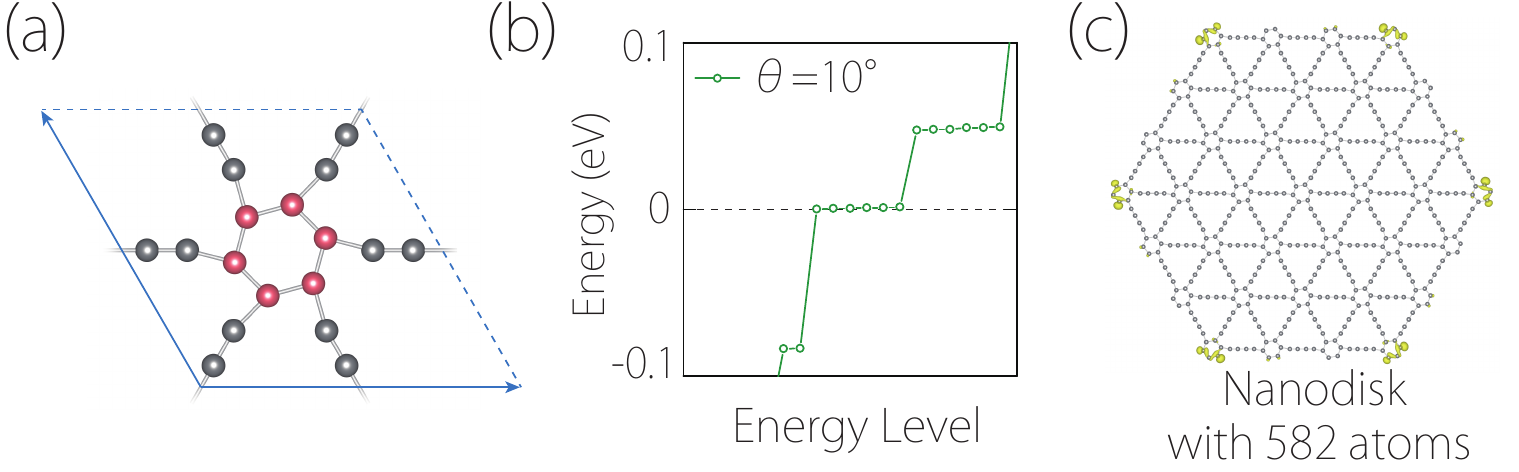}
\caption{(a) shows an artificial distortion that breaks all vertical mirrors in GDY.
Here, the benzene ring (red) in a unit cell is rotated by $10^{\circ}$. (b) shows the corresponding calculated energy spectrum. The zero-energy states persist and they are still localized near the corners, as in (c).}
\label{rotate}
\end{figure}

The above analysis have clarified the topological origin of the corner states,
which in turn proves that GDY is a SOTI.
We have used the crystalline symmetries such as $\mathcal{M}_y$ to facilitate the argument. {However, they are not \emph{required} for the existence of the topological 0D states~\cite{Langbehn2017}. This is evident from the edge picture: As long as the bulk and edge gaps are not closed, the topological classification for each edge cannot be changed by any symmetry-breaking perturbations.} For example, Fig.~\ref{rotate}(a) shows a distortion which breaks all vertical mirrors. The result in Fig.~\ref{rotate}(b,c) clearly demonstrates the corner states are indeed robust.

{\color{blue}{\em Discussion.}}---We have revealed GDY as the first realistic example of a 2D SOTI, with topological 0D corner states. In experiment, the corner state can be detected as sharp peaks in the scanning tunneling spectroscopy (STS) measurement, which does not appear in the bulk but emerges when the tip moves close to the corner.

We have shown that the corner states are robust against crystal-symmetry-breaking perturbations, which greatly widens their experimental relevance. The argument also applies to nanodisk geometry~\cite{Langbehn2017,SongZD2017}, i.e., the corner states should exist also for nanodisks with other shapes not restricted to a hexagon. Fortunately, we note that
hexagon-shaped GDY nanodisks as studied in this work naturally result from bottom-up synthesis approaches~\cite{MatsuokaJACS},
and techniques for making atomically sharp and clean edges have been developed for graphene~\cite{Jia_Sci2009,Girit1705,Rizzo2018vk} (hopefully also for GDY in near future). All these facilitate the experiment on GDY.

The corner states are pinned in the bulk gap by the approximate chiral symmetry of GDY, which is advantageous for their detection. (While chiral symmetry is approximate for insulators, it can be exact in superconductors protecting true zero-energy corner modes~\cite{ZhangF2018,YanZB2018}.) If the nanodisk is too small, the corner states may interact with each other and get repelled from zero energy. Nevertheless, our calculation shows that a nanodisk with width $\sim 6$ nm is already enough to suppress the coupling. Meanwhile, the edge modification and adsorption may strongly affect the boundary spectrum. They may locally break the chiral symmetry and shift the corner states out of the bulk gap. Thus, the edge contamination should be avoided in experiment as much as possible.

\textit{Acknowledgments---} The authors thank D. L. Deng for helpful discussions. This work is supported by the Singapore MOE AcRF Tier 2 (MOE2017-T2-2-108), the NSFC (Grants No. 11834014, No. 11504013, No. 11874201), the Fundamental Research Funds for the Central Universities (Grant No. 0204/ 14380119), and the GRF of Hong Kong (Grant No. HKU173057/17P).

 X.-L. Sheng and C. Chen contributed equally to this work.

\bibliographystyle{apsrev4-1}{}
\bibliography{GDY_ref}



\end{document}